\documentclass[12pt,preprint]{aastex}
\usepackage{natbib}

\newcommand{\etal}{et al.}
\newcommand{\msun}{M$_{\sun}$\,}

\shorttitle{Limits on the Near-IR Variability of Sgr A*}
\shortauthors{Hornstein, Ghez, Tanner, Morris, Becklin, \& Wizinowich}

\begin{document}

\title{Limits on the Short Term Variability of Sagittarius A* in the Near-Infrared}

\author{S.D. Hornstein\altaffilmark{1},
  A.M. Ghez\altaffilmark{1,2}, A. Tanner\altaffilmark{1},
  M. Morris\altaffilmark{1}, E.E. Becklin\altaffilmark{1},
  P. Wizinowich\altaffilmark{3}} 
\email{seth@astro.ucla.edu}

\begin{abstract}
The recent detection of a 3-hr X-ray flare by the Chandra
Observatory has raised the possibility of enhanced emission
over a broad range of wavelengths from Sgr A*, the suspected
2.6 x 10$^{6}$ \msun black hole at the Galactic Center,
during a flaring event. We have, therefore, reconstructed 3-hr
data sets from 2\micron ~speckle and adaptive optics images
($\theta_{core}$ = 50 - 100 mas) obtained with the W. M. Keck
10-m telescopes between 1995 and 2001. In 25 separate
observations, no evidence of any significant excess emission
associated with Sgr A* was detected. The lowest of our
detection limits gives an observed limit for the quiescent
state of Sgr A* of 0.09$\pm$0.005 mJy, or, equivalently, a
dereddened value of 2.0$\pm$0.1 mJy, which is a factor of 2
lower than the best previously published quiescent
value. Under the assumption that there are random 3-hr flares
producing both enhanced X-ray and near-infrared emission, our
highest limit constrains the variable state of Sgr A* to
$\lesssim$0.8 mJy (observed) or 19 mJy (dereddened).  These
results suggest that the model favored by \citet{mark01}, in
which the flare is produced through local heating of
relativistic particles surrounding Sgr A* (e.g., a sudden
magnetic reconnection event), is unlikely, because it predicts
peak 2\micron ~emission of $\sim$300 mJy, well above our
detection limit.
\end{abstract}

\keywords{Galaxy: center --- infrared: galaxies --- X-rays:
galaxies --- accretion, accretion disks --- galaxies: jets ---
black hole physics}

\altaffiltext{1}{Department of Physics and Astronomy,
University of California, Los Angeles, CA 90095-1562}
\altaffiltext{2}{Institute for
Geophysics and Planetary Physics, University of California,
Los Angeles, CA 90095-1565}
\altaffiltext{3}{W.M. Keck
Observatory, 65-1120 Mamalahoa Hwy., Kamuela, HI 96743}

\section{Introduction} \label{intro}

The variability of Sagittarius (Sgr) A* at X-ray wavelengths
\citep{bag01a} has bolstered the case for associating this
source with the suspected 2.6 x 10$^{6}$ \msun black hole at
the center of our Galaxy \citep{eckart97, genz97, genz00,
ghez98, ghez00}. In two Chandra observations separated by
almost a year and having a total of 76 ksec of exposure time,
Sgr A* was detected at X-ray wavelengths for the first time
and was also seen to flare in intensity over a time scale of 3
hours \citep{bag01a,bag01b}. While the flare's short duration
implied a small region of origin, $\lesssim$400 R$_{s}$ (where
$R_{s}$ is the Schwarzschild radius =
2GM$_{\bullet}$/c$^{2}$), its large amplitude, a factor of 50,
has raised the possibility of detecting corresponding
intensity enhancements at wavelengths outside the X-ray
regime.

Existing models for Sgr A*'s flared state make very disparate
predictions for the emission at wavelengths between the X-ray
and radio regimes \citep{mark01,liu02}. The wide differences
between these models are a result of assuming different
geometries (disk vs. jet) and emission mechanisms for the
flaring process (e.g., enhanced accretion rates vs. magnetic
reconnection). In some models, the predicted emission in the
flared state, at infrared (IR) wavelengths, dramatically
exceeds that of existing detection limits \citep{genz99,
stol99, morris01}. For example, the preferred model of
\citet{mark01} predicts an observed 2\micron ~flux density of
$\sim$13 mJy, or a dereddened flux density of $\sim$300 mJy,
during the flared state. Unlike the situation at radio
wavelengths, where long-term monitoring campaigns have been
used to constrain the flared state of Sgr A* \citep{zhao01},
the limited time coverage and spatial resolution of published
IR experiments prevent meaningful constraints on the flared
state's IR emission from being inferred\footnote{While several
papers have reported the possible detection of a variable
near-infrared source coincident with Sgr A*
\citep{herbst93,close95,genz97}, subsequent high resolution
observations have identified this emission to be from high
proper motion sources \citep{eckart95, eckart97, ghez98,
ghez02}.} and, thus, the reported limits are assumed to be
associated with Sgr A*'s quiescent state.

The W. M. Keck Observatory dynamical study of stars in the
central stellar cluster \citep{ghez98,ghez00,suvi02} provides
a rich source of high angular resolution 2\micron ~data
between 1995 and 2001. In this paper, we present 2\micron
~flux density limits from maps that were each composed of data
from a single night. The elapsed time of 3-4 hours in each map
is approximately the same as the time scale of the observed
X-ray flare, making this data set ideally suited for possibly
detecting a flare of this type. Given the quantity of these
observations, our non-detections establish a robust upper
limit for the flared state's 2\micron ~emission intensity.

\section{Observations}\label{obs}
High resolution, near-infrared observations of the Galactic
Center were conducted from 1995 June to 2001 July using both
speckle and adaptive optics (AO) imaging techniques on the
Keck 10-m telescopes. The speckle observations were obtained
in the K-band ($\lambda_{o}$ = 2.2\micron,
$\Delta\lambda$=0.4\micron) using the Keck I facility
near-infrared camera (NIRC; \citealt{matt94}) with external
re-imaging optics. This resulted in a pixel scale of
0\farcs0203 and a field of view (FOV) of
5\farcs12$\times$5\farcs12 \citep{matt96}. During each night
of observations several thousand short exposures (t$_{exp}$=
0.137 sec) were taken in sets of $\sim$200.  A more limited
set of data was collected using two different science cameras
behind the Keck II AO system \citep{wiz00b}. The first AO data
set was collected in the K'-band ($\lambda_{o}$ = 2.1\micron,
$\Delta\lambda$=0.35\micron) in early 1999 with the
near-infrared engineering camera (KCAM; \citealt{wiz00a}), which
had a pixel scale of 0\farcs0175, a FOV of
4\farcs4$\times$4\farcs4. Each image had an exposure time of 5
sec. The slit-viewing camera of NIRSPEC (SCAM;
\citealt{mclean98}) provided a second set of AO images for
this study. These images, like the speckle images, were made
in the K-band and have a pixel scale of 0\farcs0170 and a FOV
of 4\farcs4$\times$4\farcs4, and exposure time of 10 sec. USNO
0600-28579500 served as the natural guide star for all of
these AO observations. Since this guide star is both faint (R
= 13.2) and distant from the target (r$\sim$30\arcsec), the AO
performance was non-optimal. Table \ref{datesobs} provides a
summary of all observations.

\section{Data Analysis \& Results}
Three basic steps constitute the data analysis process in this
program. First, high angular resolution maps are generated
from the individual short exposure frames
(\S\ref{construction}). Second, all point sources in the FOV
are identified and a direct detection of Sgr A* is ruled out
(\S\ref{id}). Third, limits for Sgr A* are derived from the
residual maps, in which all identified point sources have been
removed (\S\ref{limits}).

\subsection{Construction of Images}\label{construction}
Image processing proceeds similarly to that carried out for
the dynamical experiment with one exception. Rather than
combining all the data from the duration of an observing run,
typically 2-3 nights, we synthesize the data over each night
to produce 27 maps, each of which is limited to an elapsed
time of 3-4 hours. Since the details of this method are
described elsewhere \citep{ghez98}, only a brief summary is
provided here.  Standard image reduction techniques are
applied to all the individual speckle and AO frames. For the
speckle data, a two stage shift-and-add (SAA;
\citealt{christ91,ghez98}) analysis then produces the final
high resolution maps. In the first stage, the 200 frames in
each set are combined to form an intermediate SAA image. Then,
these multiple intermediate SAA images (from throughout the
night) are combined to form one final SAA map. This allows
each intermediate image to be examined for seeing quality. In
combining the intermediate images, a seeing cut is established
so as to exclude those images with the worst seeing from the
final map. For the AO data, this cut is also carried out on
the individual AO images before they are registered and
averaged together. Figure \ref{maps} displays representative
final nightly speckle and AO maps.

\subsection{Point Source Identification \& Search for Sgr A*'s near-infrared emission}\label{id}
In all maps, stars are identified using StarFinder, an IDL
package developed for astrometry and photometry in crowded
stellar fields \citep{diolaiti00}.  This package iteratively
generates estimates of the point spread function (PSF) from a
few selected bright stars and then identifies point sources
over the entire FOV through cross-correlation of the map with
the PSF model. For the PSF extraction, we found that the most
reliable PSF models are obtained with a support size of
$\sim$2\arcsec, which represents a compromise between needing
to accommodate the large PSF halos and yet having a limited
FOV. This choice limits our analysis to images with PSF halo
sizes of 0\farcs4 or less, as the PSFs of the remaining two
images are poorly characterized by this process. The PSF model
is based on four of the five brightest stars in the FOV (IRS
16NE, 16C, 16NW, and 29N; see Figure \ref{maps}); IRS 16SW is
avoided as a PSF model star, since it is surrounded by
relatively bright stars in its immediate vicinity. For point
source identification, a correlation coefficient greater than
0.8 between the PSF model and the actual stellar image is
required to avoid spurious detections. This process results in
the identification of $\sim$100 point sources in each map.

The speckle and AO images have significantly different
PSFs. Nonetheless, both PSFs are composed of a compact core on
top of a broader halo. Table \ref{datesobs} provides the
characterization of the PSF in each map based on the radial
profile of the PSF model.  While the speckle images have PSF
core FWHM that are nearly diffraction limited
($\sim$0\farcs05) and $\sim$40\% smaller than that of the AO
images ($\sim$0\farcs08), the AO PSF core contains $\sim$30\%
of the total energy, $\sim$12 times more than the typical
speckle PSF.

At this stage in the analysis, it is possible to look for
direct detections of Sgr A*. We use proper motion acceleration
vectors \citep{ghez00} to pinpoint the central black hole's
position relative to the nominal radio position of Sgr A*
\citep{menten97} to within 0\farcs04 (1$\sigma$). Within
0\farcs08 of this location, 4 sources (13.9~$<$~K~$<$~16.5)
are identified, all of which were previously detected in
"monthly" maps made from all data in a single observing run
\citep{ghez98,ghez02} and, furthermore, have significant
proper motions.  This high stellar density emphasizes the need
for improved accuracy in Sgr A*'s position in the IR reference
frame in order to measure or constrain its emission. With no
stationary source identified in this region, we conclude that
Sgr A* has not been detected.

\subsection{Flux Density Limits for Sgr A*}\label{limits}
In order to determine an accurate detection limit at the
position of Sgr A*, it is necessary to remove the
contaminating seeing halos from nearby sources.  A `stars
only' map is created using the PSF model and list of stars
generated by StarFinder. This is then subtracted from the
original map, producing a residual map.  With the residual
map, a 3$\sigma$ point source detection limit for Sgr A* is
established based on three times the RMS of 25 aperture
photometry values, which are calculated using $\sim$60 mas
radius apertures and sky annuli extending from $\sim$60 mas to
$\sim$90 mas. The 5$\times$5 grid of apertures in the residual
map corresponds to an area of $\sim$0\farcs6$\times$0\farcs6,
more than two orders of magnitude larger than the uncertainty
in the location of Sgr A*.

Zero points are obtained through carrying out the same
aperture photometry in the original maps (prior to the `stars
only' subtraction) on all known non-variable sources brighter
than K=10.5,using the flux densities reported in
\citet{blum96}, and that occur in more than 30\% of the frames
for each night. The photometric calibration sources used are
IRS 16NW, 16C, 16CC, and 16NE, when the FOV allows its
inclusion; IRS 29N is omitted as it is found to be marginally
variable at the 2$\sigma$ level. Typical photometric zero
point 1$\sigma$ uncertainties of $\sim$0.04 mag result from
this procedure.

Table \ref{datesobs} and Figure \ref{magvstime} contain the
resulting 3$\sigma$ point source detection limits for Sgr A*.
The lowest of these upper limits gives an observed limit for
the quiescent state of Sgr A* of 0.09$\pm$0.005 mJy, or,
equivalently, a dereddened value of 2.0$\pm$0.1 mJy, while the
highest limit constrains our analysis of the variable state of
Sgr A* to $\lesssim$0.8 mJy (observed) or 19 mJy (dereddened).
Although these upper limits vary significantly from epoch to
epoch (due to variable observing conditions and/or a variation
in the number of contaminating sources detected and removed)
the lowest of them is lower than the best previously reported
limits at 2\micron~ for Sgr A*'s quiescent state (dereddened 4
mJy; \citealt{genz99}) by a factor of 2.

\section{Discussion}\label{results} 
Using the coverage of the X-ray and near-infrared experiments
and assuming random 3-hr flaring events that produce both
enhanced X-ray and near-infrared emission, we consider the
likelihood that a near-infrared flare in excess of our weakest
limit occurred over the course of the IR experiment.  The
probability of seeing such a flare is given by the following
binomial distribution:
\begin{equation} \label{binomial}
P(\textrm{$\nu$ detections in $n$ trials}) =
\frac{n!}{\nu!(n-\nu)!}p^{\nu}q^{n-\nu}
\end{equation}
where $p$ is the probability of detecting a flare in one trial
and $q$ is the probability of no detection (1-$p$).  Here we
will refer to $p$ as the duty cycle and note that in our
context, this quantity describes what fraction of a set of
3-hr images should contain a flare. Figure \ref{duty_cycle}
shows the probability distributions as a function of
underlying duty cycle for both the X-ray experiment, in which
$\nu$=1 and $n$=7.6 (obtained by dividing the total time, 76
ksec, by the length of the observed X-ray flare, 10 ksec) and
the near-infrared experiment, in which $\nu$=0 and n=25. The
X-ray and near-infrared probability distributions have only a
small overlap, which suggests that if the flaring activity at
X-ray and near-infrared wavelengths are coupled, a flare most
likely occurred during the near-infrared experiment.  The
joint probability distribution quantifies this and suggests
there is, at most, a probability of 9\% of a random 3-hr
near-infrared flare in excess of 19 mJy (dereddened). We
therefore assume that at the 2$\sigma$ confidence level, a
flare occurred during our experiment and use our limits to
constrain the variability models.

While only a limited amount of modeling of the recent 3-hr
X-ray flare detected at Sgr A* has been carried out, existing
models predict 2\micron ~dereddened emission as high as 300
mJy in the model preferred by \citet{mark01} but as small as
0.4 mJy in \citet{liu02}. The former model explains the
elevated X-ray emission, produced by synchrotron self-Compton,
by an enhanced temperature for the relativistic electron
population, as might arise in a magnetic reconnection
event. On the other hand, Liu \& Melia present a flare model
in which the flare arises due to an enhanced accretion rate
and bremsstrahlung emission is dominant at both near-infrared
and X-ray wavelengths.  The lack of a near-infrared detection
of Sgr A* makes the \citet{mark01} model and any other
mechanism that produces flared 2\micron ~emission in excess of
19 mJy (dereddened) unlikely.

\section{Conclusions}\label{conclus}
This paper summarizes a search for a near-infrared counterpart
to Sgr A* in the flared state. From the length of our
observations, this search was sensitive to variability on time
scales of 3 hours. No such counterpart was detected. However,
by identifying and removing all the stars in the crowded inner
$\sim$0\farcs6$\times$0\farcs6 of the Galactic Center, an
upper limit for the emission from Sgr A* has been inferred for
each observation epoch. These limits constrain the quiescent
emission from Sgr A* to $\lesssim$0.09 mJy (2.0 mJy,
dereddened) and the variable component to $\lesssim$0.8 mJy
(19 mJy, dereddened) at the 2$\sigma$ confidence level.  More
X-ray data will improve the estimated flaring duty cycle and
is likely to increase the confidence level of our limits for
the near-infrared component of the flared emission.

\acknowledgments Valuable input on the manuscript was
contributed by Gaspard Duch\^{e}ne, Caer McCabe, and Andreas
Eckart. Data presented herein were obtained at the W.M. Keck
Observatory, which is operated as a scientific partnership
among the California Institute of Technology, the University
of California and the National Aeronautics and Space
Administration. The Observatory was made possible by the
generous financial support of the W.M. Keck Foundation.  This
work was supported by the NSF both through grant No. 9988397
and the Science and Technology Center for Adaptive Optics,
managed by the University of California at Santa Cruz under
cooperative agreement No. AST - 9876783. AMG also thanks the
Packard Foundation for its support.

\begin{deluxetable}{lccccccc}
\tablenum{1}
\label{datesobs}
\tabletypesize{\scriptsize}
\tablecaption{Near-Infrared Limits on Sagittarius A*}
\tablewidth{0pt}
\tablehead{
\colhead{}  &  \colhead{} & \colhead{}  & \multicolumn{3}{c}{PSF
  Characteristics}  & \colhead{} &\colhead{} \\ 
\cline{4-6}
\colhead{Epoch (UT)} &
\colhead{Camera} &
\colhead{\# frames} &
\colhead{FWHM$_{core}$ (\arcsec)} &
\colhead{FWHM$_{halo}$ (\arcsec)} &
\colhead{E$_{core}$ (\%)\tablenotemark{a}} &
\colhead{Limit (mag)\tablenotemark{b}} &
\colhead{Limit (mJy)\tablenotemark{b}}
}
\startdata
1995 Jun 11  &  NIRC   &  2716 & 0.05 & 0.3 & 3 & 15.87  &  6.5\\
1996 Jun 25  &  NIRC   &  3866 & 0.06 & 0.5 & 1 & 13.71\tablenotemark{c}  & 47.5\tablenotemark{c}\\ 
1996 Jun 27  &  NIRC   &  1600 & 0.06 & 0.4 & 2 & 14.73  & 18.6\\ 
1997 May 14  &  NIRC   &  2800 & 0.05 & 0.2 & 1 & 15.67  &  7.8\\
1998 Apr 2   &  NIRC   &  2744 & 0.05 & 0.4 & 2 & 15.01  & 14.3\\
1998 May 14  &  NIRC   &  3436 & 0.05 & 0.4 & 1 & 14.79  & 17.5\\
1998 May 15  &  NIRC   &  6468 & 0.05 & 0.3 & 2 & 16.31  &  4.3\\
1998 Aug 4   &  NIRC   &  4241 & 0.05 & 0.4 & 2 & 15.33  & 10.7\\
1998 Aug 5   &  NIRC   &  6272 & 0.05 & 0.3 & 3 & 16.62  &  3.3\\
1999 May 2   &  NIRC   &  7722 & 0.05 & 0.2 & 2 & 15.96  &  6.0\\
1999 May 3   &  NIRC   &  1764 & 0.05 & 0.2 & 2 & 15.85  &  6.6\\
1999 May 4   &  NIRC   &  1960 & 0.05 & 0.3 & 2 & 15.36  & 10.4\\
1999 May 27  &  KCAM+AO&    15 & 0.08 & 0.2 &38 & 17.02  &  2.6\\
1999 Jun 28  &  KCAM+AO&    58 & 0.08 & 0.2 &41 & 16.94  &  2.8\\
1999 Jul 24  &  NIRC   &  5677 & 0.05 & 0.2 & 4 & 17.17  &  2.0\\
1999 Jul 24  &  KCAM+AO&    43 & 0.11 & 0.2 &35 & 16.33  &  4.9\\
2000 Apr 21  &  NIRC   &  2744 & 0.05 & 0.5 & 1 & 14.20\tablenotemark{c}  & 30.4\tablenotemark{c}\\
2000 May 19  &  NIRC   &  8232 & 0.05 & 0.3 & 3 & 16.43  &  3.9\\
2000 May 20  &  NIRC   &  6860 & 0.05 & 0.2 & 4 & 16.53  &  3.5\\
2000 Jun 21  &  SCAM+AO&   215 & 0.08 & 0.2 &26 & 15.41  &  9.9\\
2000 Jun 22  &  SCAM+AO&    54 & 0.08 & 0.2 &24 & 16.84  &  2.7\\
2000 Jul 19  &  NIRC   &  5473 & 0.06 & 0.3 & 4 & 16.24  &  4.6\\
2000 Jul 20  &  NIRC   &  3255 & 0.06 & 0.3 & 5 & 16.35  &  4.2\\
2000 Oct 18  &  NIRC   &  2286 & 0.05 & 0.3 & 2 & 15.73  &  7.4\\
2001 May 9   &  NIRC   &  6427 & 0.05 & 0.3 & 3 & 17.03  &  2.2\\
2001 Jul 28  &  NIRC   &  5684 & 0.05 & 0.2 & 5 & 16.61  &  3.3\\
2001 Jul 29  &  NIRC   &  3920 & 0.05 & 0.3 & 3 & 16.04  &  5.5\\
\enddata

\tablenotetext{a}{Percentage of total energy contained in the
  PSF core}

\tablenotetext{b}{Column 7 lists observed limits and column 8
  lists dereddened values using an A$_{v}$=30 with
  A$_{k}$/A$_{v}$=0.112 and A$_{k'}$/A$_{v}$=0.117
  \citep{melia01, rieke85}}

\tablenotetext{c}{Limits for the two images with halo FWHM
  $\geq$0\farcs5 are not included in the analysis (see
  \S\ref{id}).}
\end{deluxetable}
\clearpage

\newpage
\figcaption[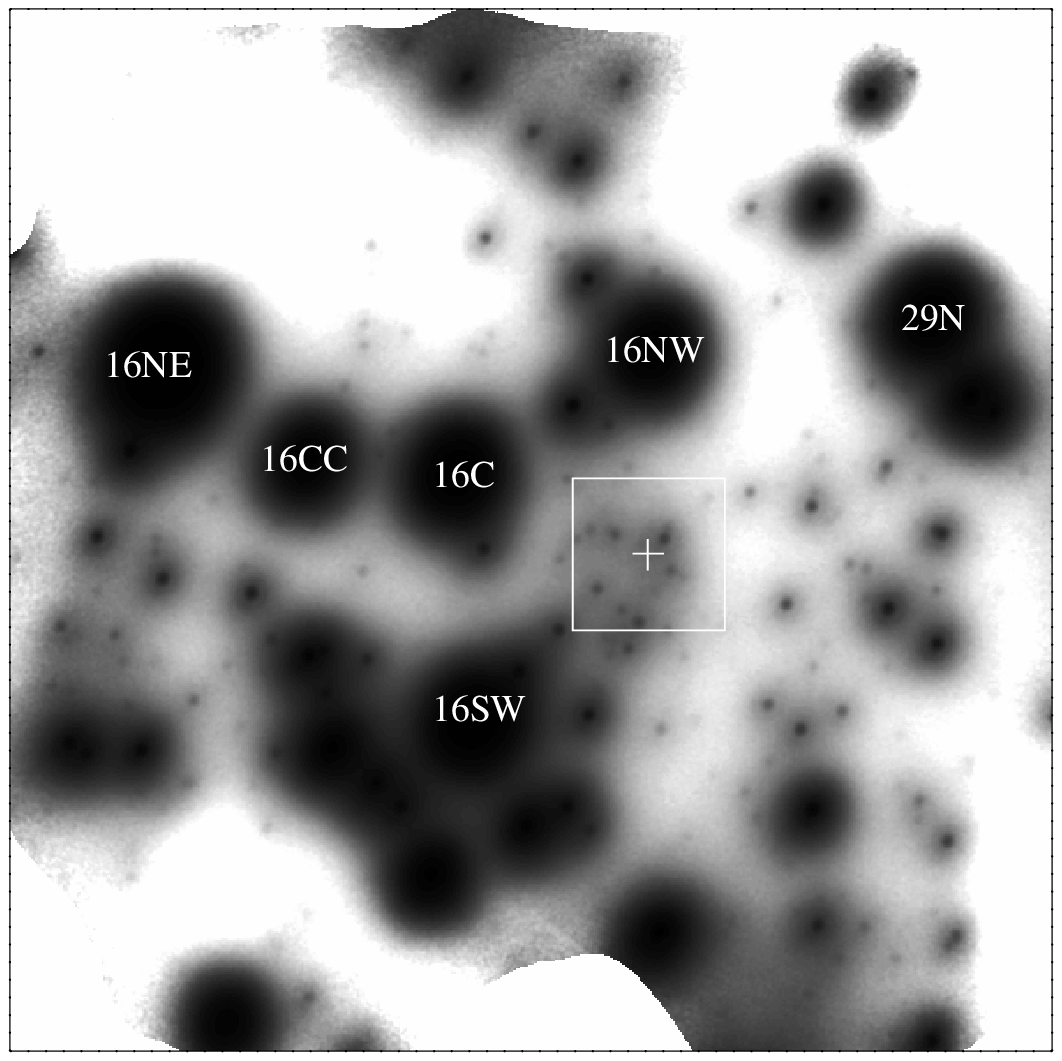]{Speckle (left) and adaptive optics (right)
    images from May 1999 of the Galactic Center. The small box
    indicates a 1\arcsec$\times$1\arcsec ~region centered on
    Sgr A*, whose approximate position is marked with a cross.
    Both images are displayed with a histogram equalization
    stretch to show the fainter stars in the field and are
    oriented such that North is up and East is to the
    left.\label{maps}}

\figcaption[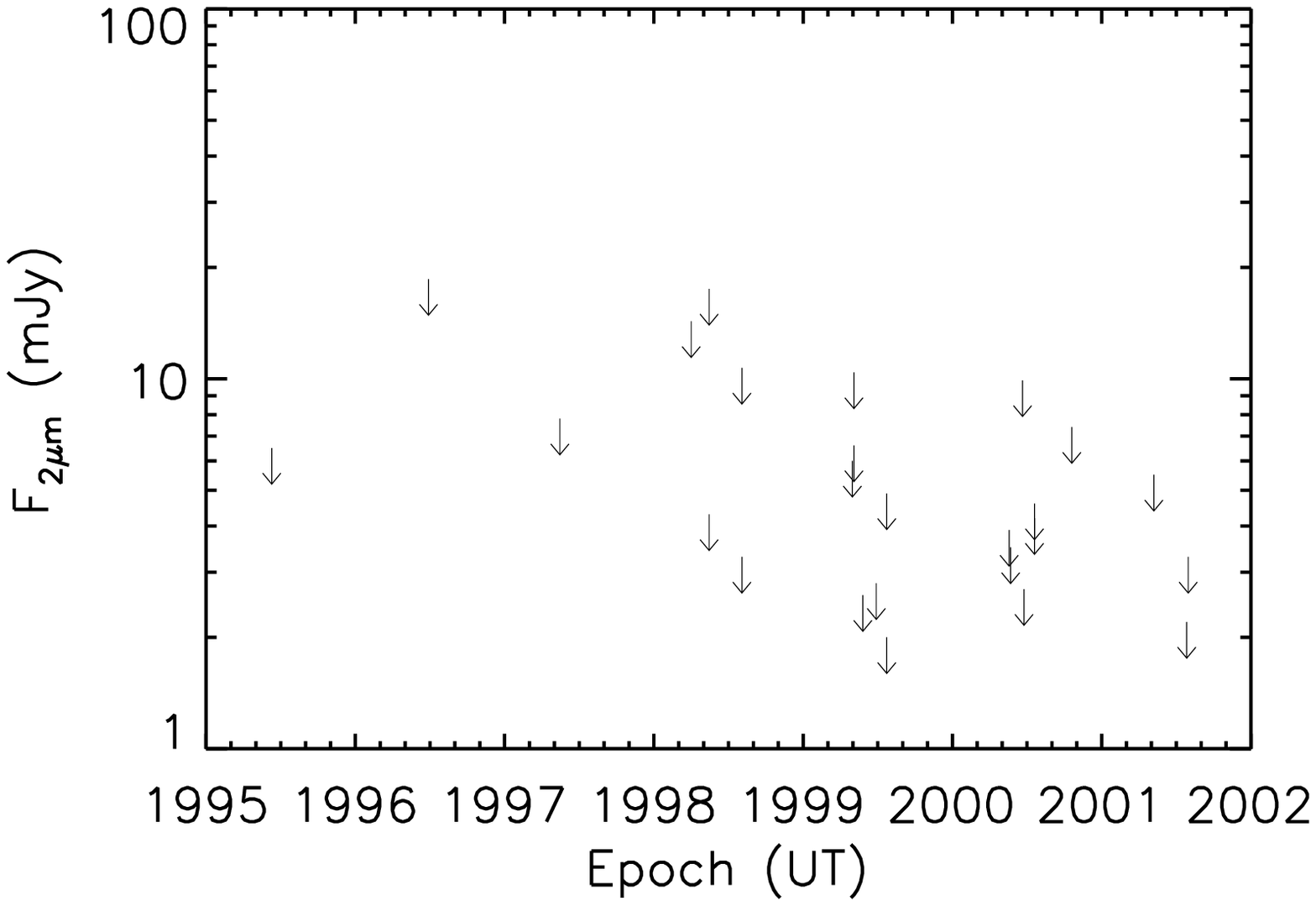]{The 3$\sigma$ limiting flux density
    calculated for Sgr A* for each epoch of observation
    (corrected for reddening by a factor of $\sim$22.) The
    lowest limit of 2.0 mJy, a factor of two lower than
    previously published limits, and the highest limit of 19
    mJy, constrains Sgr A*'s quiescent and variable states,
    respectively.\label{magvstime}}

\figcaption[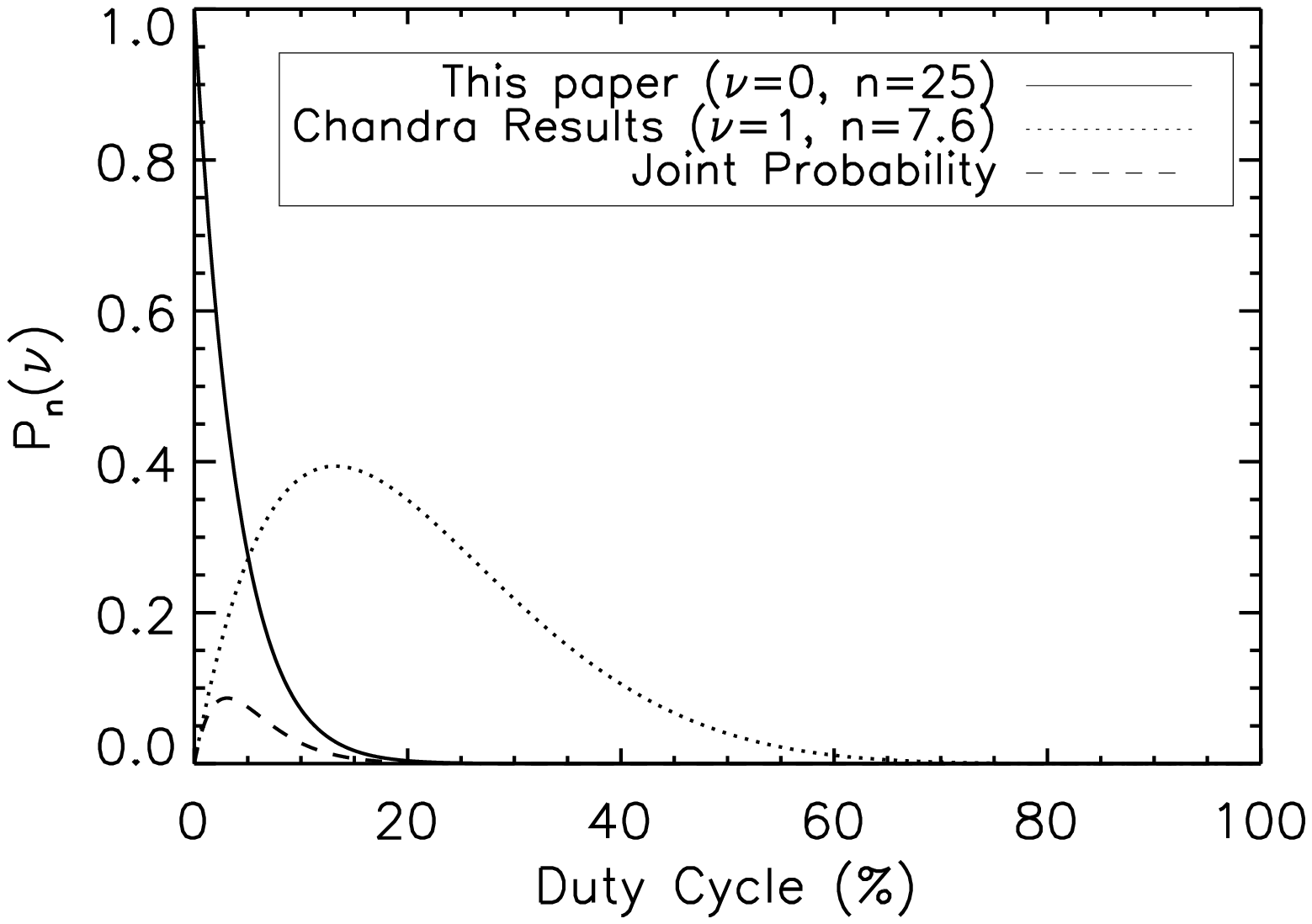]{The probability (P$_{n}$) of seeing the
    number of detected flares ($\nu$) of both this study (0
    detections) and the Chandra experiment (1 detection;
    \citealt{bag01a}) for various duty cycles. The low joint
    probability suggests that if the flaring activity at
    near-infrared and X-ray wavelengths are coupled, a flare
    occurred during the near-infrared experiment and is
    restricted to be lower than our detection limit.
\label{duty_cycle}}

\clearpage
\begin{figure*}
\plottwo{f1.eps}{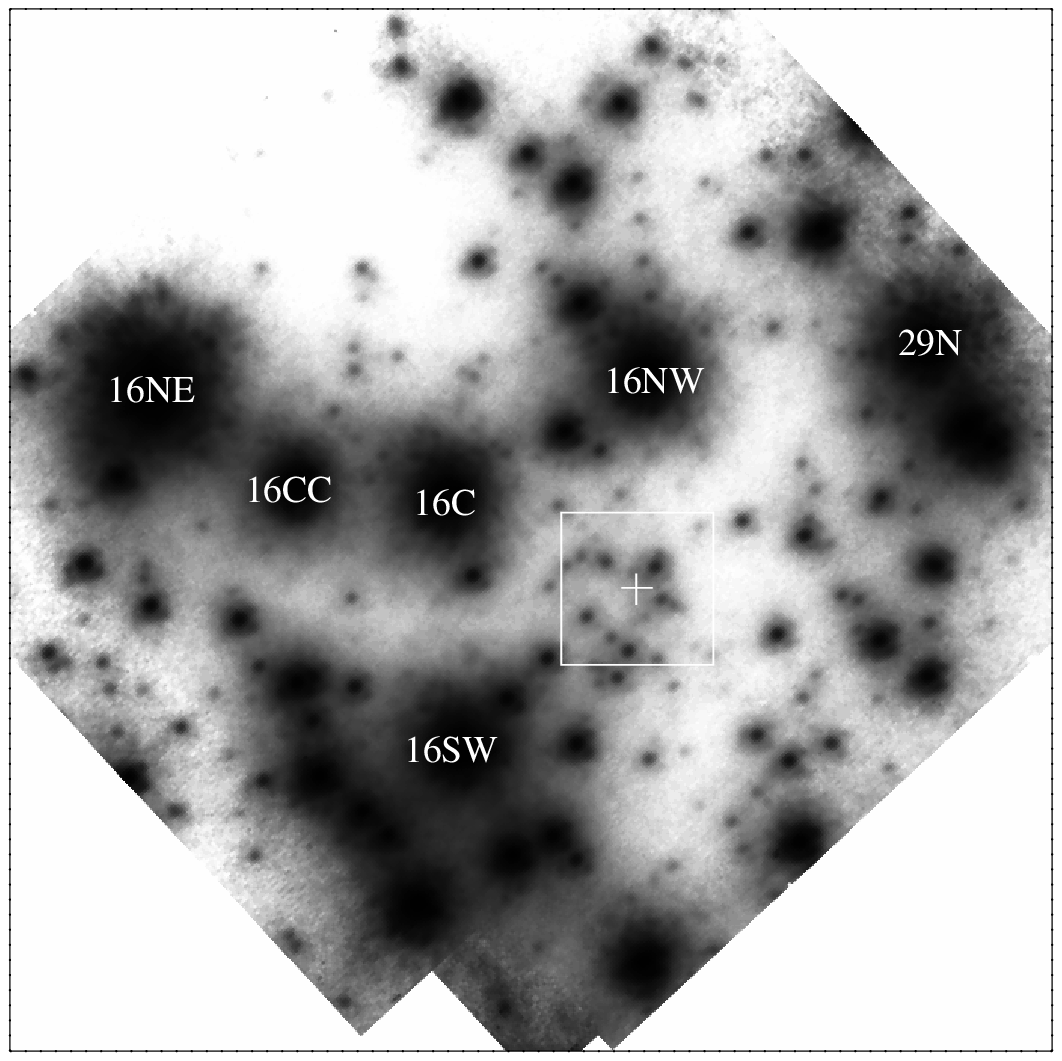}
\end{figure*}

\begin{figure}
\plotone{f4.eps}
\end{figure}

\begin{figure}
\plotone{f5.eps}
\end{figure}
\end{document}